# A Novel Low Power UWB Cascode SiGe BiCMOS LNA with Current Reuse and Zero-Pole Cancellation

Chunbao Ding, Wanrong Zhang[1], Dongyue Jin, Hongyun Xie, Pei Shen, Liang Chen,

School of Electronic Information and Control Engineering, Beijing University of Technology,

Beijing 100124, China

*Abstract*—A low power cascode SiGe BiCMOS low noise amplifier (LNA) with current reuse and zero-pole cancellation is presented for ultra-wideband (UWB) application. The LNA is composed of cascode input stage and common emitter (CE) output stage with dual loop feedbacks. The novel cascode-CE current reuse topology replaces the traditional two stages topology so as to obtain low power consumption. The emitter degenerative inductor in input stage is adopted to achieve good input impedance matching and noise performance. The two poles are introduced by the emitter inductor, which will degrade the gain performance, are cancelled by the dual loop feedbacks of the resistance-inductor (*RL*) shunt-shunt feedback and resistance-capacitor (*RC*) series-series feedback in the output stage. Meanwhile, output impedance matching is also achieved. Based on TSMC 0.35μm SiGe BiCMOS process, the topology and chip layout of the proposed LNA are designed and post-simulated. The LNA achieves the noise figure of 2.3~4.1dB, gain of 18.9~20.2dB, gain flatness of ±0.65dB, input third order intercept point (*IIP*$_3$) of -7dBm at 6GHz, exhibits less than 16ps of group delay variation, good input and output impedances matching, and unconditionally stable over the whole band. The power consuming is only 18mW.

## I. INTRODUCTION

The standard of Ultra-Wideband (UWB) was set up and approved 7.5 GHz band (3.1-10.6GHz) for UWB applications by Federal Communications Commission (FCC) in 2002. UWB receivers have some advantages such as strong anti-interference, high transmission rate, wide frequency bandwidth, and low cost[1]. As the first stage of UWB receivers, low noise amplifier (LNA) performance has an important influence on the whole receiver system. UWB LNA should have low noise figure and high gain to elevate receiver signal to noise ratio (SNR), low power consumption to preserve battery powered life of portable devices, good impedance matching to reduce return loss, and absolute stability in the whole band.

To obtain high gain, distributed and multistage LNAs usually cascade several stages [2,3]. For multistage topologies, multiple dc bias paths are needed, which largely increase the total power consumption. To realize low power, common gate (CG)-common source (CS) and CS-CS current-reuse structures[4][5][6] are generally adopted, meanwhile, high gain also can be achieved. Therefore in this paper, the current reuse topology is used, but in a novel cascode-common emitter (CE) topology style. In the topology, the cascode structure is used as the input stage to improve the reverse isolation of the LNA. The good input impedance matching and noise performance are achieved by emitter inductive degeneration technique. In order to cancel two poles arising from the emitter inductor, and hence to flat the gain performance, the dual feedback topology of *RL* shunt-shunt feedback and *RC* series-series feedback in the output stage is proposed, meanwhile, the output impedance matching is also achieved. Finally, based on SiGe HBT technology, the UWB LNA is designed and realized.

[1] Corresponding author. Tel.: +86 10 67396131-1
E-mail address: wrzhang@bjut.edu.cn

## II. Topology and Analysis of the Proposed LNA

A    The Proposed Current-Reuse Cascode LNA

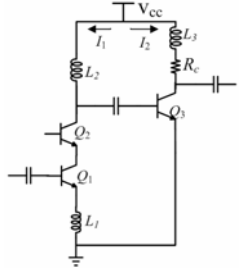 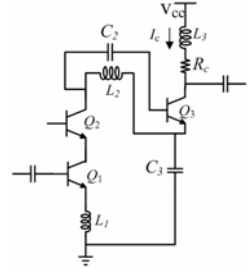

Figure 1 Schematic of traditional two stages LNA    Figure 2 Schematic of the proposed current reuse LNA

Fig. 1 is a traditional two stage LNA topology. The cascode structure is used as the input stage due to its high reverse isolation and good frequency characteristic[7], common emitter transistor $Q_3$ is the output stage for amplifying the input signal again to achieve high gain. Input and output stages are connected as a cascaded structure by capacitor coupling path. Note that two bias currents provide for cascode and $Q_3$, large power consumption is inevitable.

In order to realize low power consumption, a current reuse topology with $L_2$ and $C_2$ is proposed, as shown in Fig. 2. It has only one current biasing stream for $Q_1$, $Q_2$ and $Q_3$ in DC path, reduces current consumption by the reuse of bias current. It is noted that in the AC path, the signal is amplified by the input stage, the series inductor $L_2$ provides a high impedance path to block the signal, and the capacitor $C_2$ decouple the AC interaction between the first and second stage, therefore the signal can be again amplified by the second stage, high gain can be achieved the same as two stages cascade topology. The bypass-capacitor $C_3$ avoids the signal interference coupling back to cascode. To show that the current reuse LNA can realize low power consumption, meanwhile achieve high gain, the analysis is given as follows.

Compared Fig. 1 with Fig. 2, the power consumption of traditional amplifier and current reuse amplifier are:

$$P_{cascade} = V_{cc1} \times (I_1 + I_2)$$
$$P_{current-reuse} = V_{cc2} \times I_c \qquad (1)$$

When $V_{cc1}=V_{cc2}$, the input impedances of traditional amplifier and current reuse amplifier are all matched well in the same sizes of $Q_1$, $Q_2$ and $Q_3$, $I_1$ is approximately equal to $I_c$. Consequently, power consumption of the current-reuse amplifier is smaller than the traditional amplifier.

Since the current gain of common base transistor $Q_2$ is nearly unity, the effective transconductance of the cascode amplifier is equal to the transconductance of $Q_1$[7]. So the effective transconductances of traditional amplifier and current reuse amplifier are same, and obtained as

$$G_{cascade} = G_{current-reuse} \approx g_{m,Q_1} \times g_{m,Q_3} \qquad (2)$$

Therefore, when $I_1 \approx I_2 \approx I_C$, the current reuse amplifier reduces the power consumption without affecting the power gain.

B.    Input Impedance and Noise Analysis

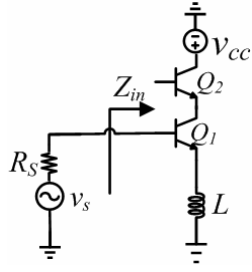
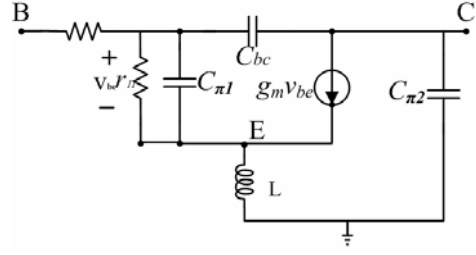

Figure 3 Input stage with emitter degenerative inductive    Figure 4 Small signal equivalent circuit

Since the LNA is the first module of the receiver, the input impedance must be matched to the source impedance (50 Ω) so as to reduce the distortion caused by signal reflection. The emitter degenerative inductive technique is adopted to achieve wideband input impedance matching, as shown in Fig. 3. Fig. 4 shows its small signal equivalent circuit, input impedance $Z_{IN}$ is derived as follows:

$$Z_{in} = r_b + \frac{1}{jwC_{\pi 1}} + jwL + \frac{L \cdot g_m}{C_{\pi 1}} \qquad (3)$$

The input impedance matching can be achieved by adjusting the inductor value and the bias current of the circuit.

The noise characteristic is also very important for the LNA. After the impedance matching is achieved, we now turn to noise characteristic analysis. According to the Friis noise figure (*NF*) equation of cascade amplifier [8], when the gain of the first stage in the LNA is high enough, the total NF of the LNA is mainly dominated by the first stage. Therefore, it is assumed that the overall noise figure mainly arises from the first stage in the following analysis. Although the common base stage in the cascode amplifier adds some noise to the LNA, the noise from it is very small at the output compared with the noise from the common-emitter stage. Thus, for simplification of calculation, the effect of common base part is omitted.

In order to calculate the noise performance, the small signal noise model and equivalent input referred noise model for cascode stage with emitter degenerative inductive technique (CAEDI) are shown in Fig. 5 and Fig. 6, respectively.

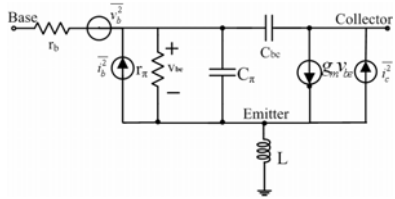
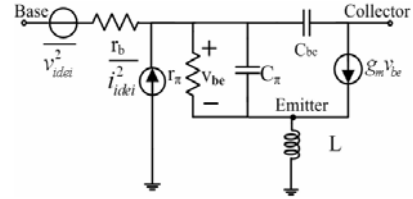

Figure 5 small signal Noise model of CAEDI    Figure 6 Equivalent input referred noise model of CAEDI

The mean square value of equivalent noise voltage $v_{iedi}$ is increased due to the emitter inductor feedback [7], can be derived as:

$$\overline{v_{iedi}^2} = 4kTr_b\Delta f + \frac{2qI_C\Delta f}{g_{mm}^2} + 2qI_B\Delta f |jwL|^2 + \frac{w^2}{w_T^2}2qI_C|jwL|^2 \Delta f \qquad (4)$$

The mean square value of equivalent noise current $i_{iedi}$ can be expressed as

$$\overline{i_{iedi}^2} = 2qI_B\Delta f + \frac{w^2}{w_T^2}2qI_C\Delta f \qquad (5)$$

where, $r_b$ is the base series resistance, $k$ is Boltzmann's constant, $q$ is the electron charge, $T$ is absolute temperature in degrees Kelvin, $\Delta f$ is frequency bandwidth of interest. $v_{iedi}$ and $i_{iedi}$ is correlated each other. The noise parameters (equivalent noise resistance $R_n$, the optimum noise impedance $Z_{OPT}$, and the minimum noise figure $NF_{min}$) are derived as follows:

$$R_n = R_b + \frac{1}{2}(\frac{g_m}{\beta} + g_m \frac{w^2}{w_T^2})w^2 L_e^2 + \frac{1}{2g_m} - \frac{w}{w_T}wL_e \tag{6}$$

$$Z_{OPT} = \frac{1}{\frac{g_m}{\beta} + \frac{w^2 g_m}{w_T^2}} \sqrt{2(\frac{g_m}{\beta} + \frac{w^2 g_m}{w_T^2})R_b + \frac{1}{\beta}} + j(\frac{w}{(\frac{w_T g_m}{\beta} + \frac{w^2 g_m}{w_T})} - wL_e)$$

$$NF_{min} = 1 + 2\sqrt{(\frac{g_m}{\beta} + \frac{w^2 g_m}{w_T^2}) + \frac{1}{\beta}}$$

The NF is calculated as:

$$NF = NF_{min} + \frac{2(\frac{g_m}{\beta} + \frac{w^2 g_m}{w_T^2})}{R_S}|Z_{OPT} - Z_S|^2 \tag{7}$$

where $g_m = qI_C/kT$, $\beta$ is the current gain, $w_T = g_m/(C_\pi + C_{bc})$, $Z_S$ is the source impedance (50 Ω). According to the Eq (6) and Eq. (7), NF can be optimized by adjusting the structure of transistors, bias current and inductor L. Therefore, in order to achieve good impedance matching and noise performance, the common emitter transistor in cascode stage is equipped with $A_E = 2\times(0.3\times10)$ μm², and the value of inductor L is 0.28 nH. Furthermore, the poles will be introduced by the emitter inductor, the gain performance is degraded. In the following we analyze that and try to compensate the gain degradation.

C. Zero-Pole Cancellation

The transfer function of cascode input stage with emitter inductive degeneration is derived as:

$$\frac{i_{out}}{v_s} = \frac{1}{(R_S + r_b + \frac{L \cdot g_m}{C_\pi}) \cdot sC_\pi + 1 + s^2 L \cdot C_\pi} \cdot \frac{g_m}{1 + g_m \cdot sL} \tag{8}$$

The term $s^2 LC_\pi + 1$ in (8) can be ignored over the band of interest since the input matching network is a low-Q circuit and the center frequency $f_o = 6.5$ GHz. Therefore, the pole $P_0$ and $P_1$ contributed by the emitter inductor are expressed as:

$$P_0 = 0$$
$$P_1 = -\frac{1}{g_m L} \tag{9}$$

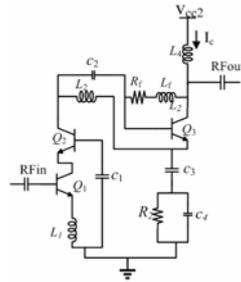

Figure 7 The proposed current reuse LNA with dual feedback output stage

The poles $P_0$ and $P_1$ degrade the gain performance of the LNA. Therefore, in order to flat the gain performance, the additional zeros should be introduced to compensate the degradation. So the resistance-inductor shunt feedback and resistance-capacitor series feedback are adopted in the output stage $Q_3$ of the LNA, as shown in Fig. 7. The transfer function of the output stage is derived as:

$$\frac{v_{out3}}{i_{in3}} = \frac{v_{out3}}{v_{in3}} \cdot z_{in3} = -(\frac{g_{m3} \cdot (1 + R_2 sC_4)}{(1 + R_2 sC_4) + g_{m3} \cdot R_2} - \frac{1}{R_f + s \cdot L_f}) \cdot \frac{(R_f + sL_f) \cdot sL_4 \cdot R_L}{(R_L + sL_4) \cdot (R_f + sL_f) + R_L \cdot sL_4}$$

$$\cdot \frac{(1 + R_2 sC_4 + g_{m3} \cdot R_2)(R_f + sL_3)[(R_L + sL_4) \cdot (R_f + sL_3) + R_L \cdot sL_4]}{[(1 + R_2 sC_4 + g_{m3} \cdot R_2) + sC_\pi (R_f + sL_f)(1 + R_2 sC_4)] \cdot [(R_L + sL_4) \cdot (R_f + sL_f) + R_L \cdot sL_4] + g_{m3} \cdot R_L \cdot sL_4 \cdot (R_f + sL_f) \cdot (1 + R_2 sC_4)} \tag{10}$$

where, $v_{out3}$, $v_{in3}$, $i_{in3}$ and $z_{in3}$ are the output voltage, input voltage, input current and input impedance of the output stage, respectively. According to the Eq. (10), three new zeros and one pole are introduced and expressed as

$$Z_0 = 0 \quad Z_1 = -\frac{g_{m3}R_2 + 2 - g_{m3}R_f}{g_{m3}L_f + (g_{m3}R_f - 2)R_2} \quad Z_2 = -\frac{R_f}{L_f} \quad (11)$$

$$P_2 = -\frac{R_f R_L (1 + g_{m3}R_2)}{R_L(1 + g_{m3}R_2)L_f + C_4(R_L R_f + R_2 R_L R_f) + R_f R_L g_{m3} L_4}$$

The pole $P_0$ from Eq. (8) can be cancelled with zero $Z_0$ from Eq. (11) introduced by the load inductor $L_4$, meanwhile pole $P_1$ introduced by the $R_f$-$L_f$ feedback is also cancelled with the zero $Z_1$ by adjusting the resistance $R_f$ and inductor $L_f$. The additional pole $P_2$ introduced by the dual feedback network will be cancelled with the zero $Z_2$ by adjusting the resistance $R_2$ and capacitance $C_4$. Therefore, the gain performance can be improved by the resistance-inductor shunt and resistance-capacitor series feedback by pole-zero cancellation.

D. Circuit Topology and Chip Layout

The complete topology of the proposed UWB LNA with current reuse is shown in Fig. 8. The input impedance matching is achieved by emitter inductor $L_1$. Output impedance matching is achieved by $R_f$-$L_3$ shunt feedback and $R_2$-$C_4$ series feedback, meanwhile the bandwidth is also extended by the pole-zero cancellation. Resistance $R_f$ and $R_1$ are used for self-biasing $Q_3$ and $Q_2$. The mirror current source Bias1 provides stable bias current for transistors $Q_1$. Inductor $L_2$ and capacitor $C_2$ are used for the current reuse structure. Based on TSMC 0.35μm SiGe BiCMOS process, the chip layout of the UWB LNA has been designed, as shown in Fig. 9, the area is 0.88×0.98 mm².

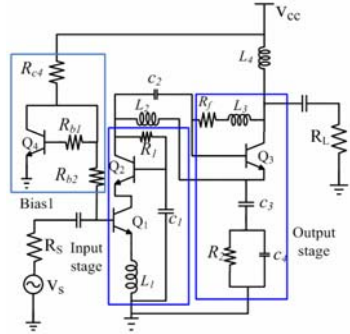
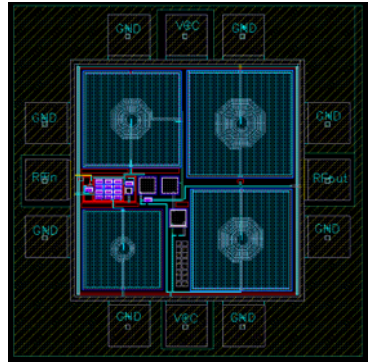

Figure 8 Topology of proposed UWB LNA      Figure 9 Chip layout of proposed UWB LNA

III Verification and Result Analysis

The proposed UWB LNA is post-simulated with Spectre of Cadence's EDA using TSMC 0.35 μm SiGe BiCMOS process design kit(PDK), the following figures show the post-simulation results.

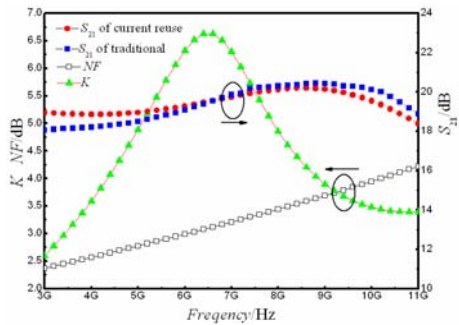
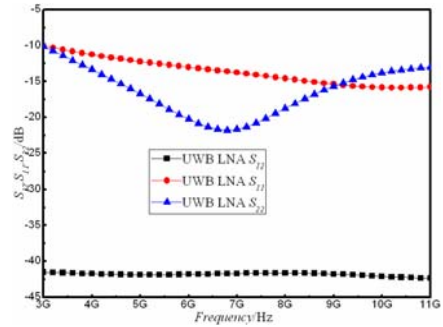

Figure 10 *NF* and $S_{21}$ and Stability Factor *K* of proposed LNA      Figure 11 $S_{11}$, $S_{22}$, $S_{12}$ of proposed LNA

Fig. 10 depicts the $S_{21}$ and NF of the proposed UWB LNA with current reuse together with $S_{21}$ of the traditional two stages LNA. The peak $S_{21}$ is 20.2 at 8.6GHz with the power consumption of 18mW. The gain flatness is ±0.65 dB from 3.1 to 10.6 GHz. The results demonstrate that the LNA with current reuse can indeed achieve the similar $S_{21}$ characteristics and the gain flatness is improved compared with cascade LNA. Nevertheless, the power consumption is approximately half of the traditional two stages LNA. The validity of the proposed current reuse and zero-pole cancellation approach is verified. In addition, the proposed UWB LNA has good noise performance, the noise figure is 2.3~4.1 dB.

The input return loss $S_{11}$, output return loss $S_{22}$ and reverse isolation $S_{12}$ versus frequency are plotted in Fig. 11. As shown, the $S_{11}$ and $S_{22}$ are all lower than -10 dB while $S_{12}$ is lower than -41 dB. The stable factor $K$ in Fig. 9 is larger than 2.6 and $|\Delta|=|S_{11}S_{22}-S_{12}S_{21}|<0.3$. All the results indicate this LNA has good impedance matching, reverse isolation and is absolute stability from 3.1 to 10.6 GHz.

As the derivation of the phase of transfer function, Group delay is usually used to evaluate phase nonlinearity. As shown in Fig.12, the group delay variation of LNA is ±16 ps, achieving good phase linearity. Fig. 13 shows the input 3$^{rd}$ order intercept point ($IIP_3$) of UWB LNA, which is -7 dBm at 6 GHz when a two tone test is performed with 10 MHz spacing for the entire UWB band. Therefore the LNA have good linearity, and meets the requirement of UWB receivers.

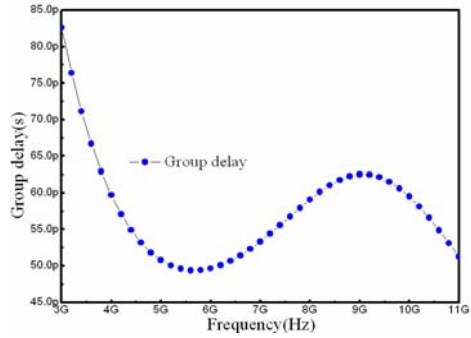 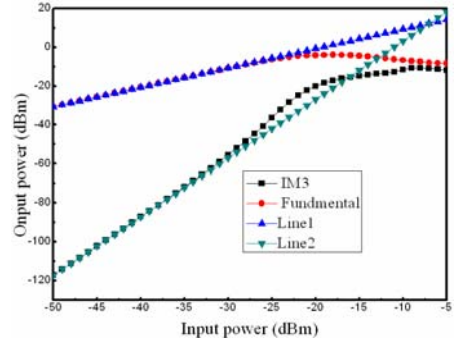

Figure12 Group delay of proposed LNA        Figure13 IIP$_3$ of proposed LNA at 6 GHz

Table I shows the summary of the proposed LNA and comparison with the recently reported SiGe and CMOS UWB LNAs. The proposed LNA has lower power consumption and better phase linearity with high gain and low noise compared with the previously published UWB LNAs [10,11,13,15]. Although the two LNAs [12, 14] have much lower power consumption, the proposed LNA exhibits better gain and group delay. In addition, the proposed LNA achieves good gain flatness.

TABLE I

SUMMARY OF THE PROPOSED SiGe UWB LNA, AND COMPARISON WITH THE RECENTLY REPORTED SiGe AND CMOS UWB LNA

| | Process | Frequency (GHz) | Peak $S_{21}$(dB) | Gain flatness(dB) | Minimum NF(dB) | Group delay variation(ps) | IIP$_3$(dBm) | Power consumption(mW) |
|---|---|---|---|---|---|---|---|---|
| This work | 0.35μm SiGe | 3.1-10.6 | 20.2 | ±0.65 | 2.3 | ±16 | -7@6GHz | 18 |
| [10] | 0.18μm SiGe | 3-10 | 20.8 | ±0.5 | 3.05 | NA | -11.7@3GHz | 42.5 |
| [11] | 0.25μm SiGe | 3.1-10.6 | 21 | ±1.25 | 2.8 | NA | -8@3GHz | 29.7 |
| [12] | 0.25μm SiGe | 4-6 | 10 | ±1.5 | 4.5 | ±30 | NA | 3.5 |
| [13] | 0.18μm CMOS | 3.1-10.6 | 17.5 | ±0.8 | 3.1 | ±48 | NA | 33.2 |
| [14] | 0.18μm CMOS | 3.1-10.6 | 9.3 | NA | 4 | ±40 | -6.7@6GHz | 9 |
| [15] | 0.18μm SiGe | 3-10 | 20.3 | ±1.05 | 1.8 | NA | 2.1@6GHz | 26 |

## IV CONCLUSION

A new topology of UWB (3.1~10.6 GHz) SiGe LNA is proposed to realize low power consumption and high gain. The LNA is composed of cascode input stage and common emitter output stage. Cascode input stage together with emitter degenerative inductive make LNA a good input impedance matching and noise performance. At output stage, the dual loop feedbacks of resistance-inductor shunt feedback and resistance-capacitor series feedback are employed to flat gain performance by the pole-zero cancellation. The proposed SiGe UWB LNA exhibits peak gain of 20.2 dB with the power consumption 18 mW, meanwhile achieves good gain flatness, low NF, better phase linearity, good input and output impedances matching over the UWB.

## REFERENCES

[1] D. Porcino and W. Hirt. Ultra-wideband radio technology: Potential and challenges ahead, IEEE Communication Magazine 2003, 41(7): 66–74.

[2] R.C. Liu, C.-S. Lin, K.-L. Deng, and H. Wang. A 0.5–14 GHz 10.6 dB CMOS cascode distributed amplifier. VLSI Circuits Symp. Tech. Dig., Jun. 2003. 139–140.

[3] Y. Shim, C.W. Kim, J. Lee, and S.-G. Lee. Design of full band UWB common-gate LNA. IEEE Microw. Wireless Compon. Lett., 2007, 17(10), 721–723.

[4] H. K. Cha, M. K. Raja, X. Y. Yuan, et al. A CMOS MedRadio Receiver RF Front-End With a Complementary Current-Reuse LNA, IEEE Transactions on Microwave Theory and Techniques, JULY 2011, 59(7), 1846-1854

[5] R. M. Weng, C. Y. Liu, P. C. Lin. A Low-Power Full-Band Low-Noise Amplifier for Ultra- Wideband Receivers. IEEE Transactions on Microwave Theory and Techniques, 2010, 58,2077-2083

[6] S. MK, Soliman AM. Low-voltage low-power CMOS RF low noise amplifier. AEU International Journal of Electronics and Communications 2009; 63(6):478–82.

[7] P. R. Gray and R. G. Meyer. Analysis and Design of Analog Integrated circuit. Fourth Edition. Beijing: Higher Education press, 2005

[8] H. T. Friis. Noise figure of radio receivers, Proc. IRE 1944; 32(7): 419–4226.

[9] Sasilatha T, Raja J. A 1 V, 2.4 GHz low power CMOS common source LNA for WSN applications. AEU International Journal of Electronics and Communications 2010, 64: 940–6.

[10] J. Lee and J. D. Cressler, Analysis and design of an ultra-wideband low-noise amplifier using resistive feedback in SiGe HBT technology, Microwave Theory and Techniques. 2006; 54(3):. 1262–1268.

[11] D. Barras, F. Ellinger, H. Jackel, and W. Hirt. A low supply voltage SiGe LNA for ultra-wideband frontends," IEEE Microw. Wireless Compon. Lett. 2004; 14(10): 469–471.

[12] B. Shi and M. Y. W. Chia. A SiGe Low-Noise amplifier for 3.1–10.6 GHz ultra-wideband wireless receivers, in IEEE Radio Freq. Integrated Circuits Symp 2006. 57–60

[13] Y. Lu, K. S. Yeo. A. Cabuk, J. Ma, M. A. Do, and Z. Lu, A novel CMOS low-noise amplifier design for 3.1-to-10.6-GHz ultra-wideband wireless receiver, IEEE Trans. Circuits Syst. I, Reg. Papers 2006; 53(8):1683–1692.

[14] A. Bevilacqua and A. M. Niknejad. An ultra wideband CMOS low noise amplifier for 3.1–10.6 GHz wireless receivers," IEEE J. Solid-State Circuit 2004: 39(12): 2259–2268.

[15] Y. Lu, R. Krithivasan, W. M. L. Kuo, and J. D. Cressler. A 1.8–3.1 dB noise figure (3–10 GHz) SiGe HBT LNA for UWB applications, in IEEE Radio Freq. Integrated Circuits Symp 2006. 45–48.